\documentclass[11pt,twoside]{article}
\usepackage{newpasp}

\index{Mashchenko, S.}
\index{Carignan, C.}

\begin{document}

\title{The Impact of External Tidal Fields on the ISM of Dwarf Spheroidals}
\author{Sergey Mashchenko and Claude Carignan}
\affil{D\'epartement de Physique and Observatoire du Mont M\'egantic,
 Universit\'e de Montr\'eal, C.P. 6128, Succursale Centre Ville,
 Montr\'eal, H3C 3J7, QC, Canada}


\begin{abstract}
It was
found that for three dwarf spheroidal galaxies  --- Sculptor, Tucana, and Cetus --- there is
a correspondence between the distribution of associated HI
gas and the projected tidal axis direction. Numerical hydrodynamical simulations confirmed the
following scenario: SN~Ia explosions remove most of the interstellar medium beyond
the tidal radius of the dwarf galaxy, with most of the mass
being in clouds moving along the principal tidal axis.
\end{abstract}




\section{Observational data}

\begin{figure}[b]
\plotfiddle{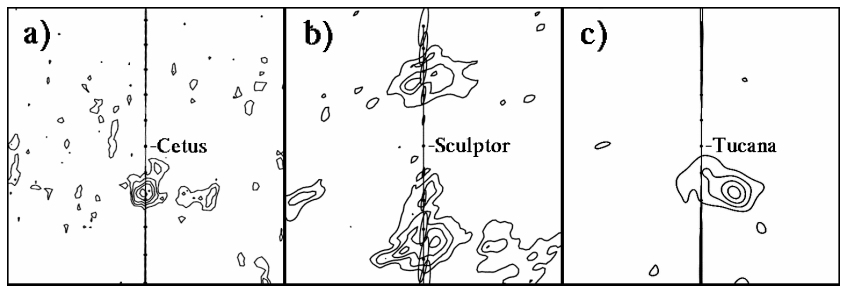}{100pt}{0}{25}{25}{-157}{0}
\caption{\small HI distribution (contours) and projection of the
 principal tidal axes (vertical lines)
 for three dwarf spheroidals}
\label{fig1}
\end{figure}

Figure~\ref{fig1} shows the HI distribution integrated over certain intervals 
of velocities for three
dwarfs spheroidal (dSph) galaxies --- Cetus, Sculptor, and Tucana.
The HI data have been obtained with the
Australia Telescope Compact Array (Sculptor and Tucana, with the Tucana data 
presented by Oosterloo, Da Costa, \& Staveley-Smith, 1996),
and the Parkes Multibeam System (Cetus).
Monte-Carlo simulated projections of the principal (``stretching'') tidal
 axis are shown as 
vertical lines.  One sigma error ellipses are shown for the apparent
location of points on the tidal axis, located at spatial distances 
$r_t$, $2r_t$, $3r_t$, \dots  from the center of the galaxy ($r_t$ is the tidal
radius.) Tidal tensor components were assumed to be constant within
the dSph, and included contributions from all Local Group members.
The images (Figure~\ref{fig1})
were rotated and rescaled to make 1) their projected
principal tidal axes vertical, and 2) their projected tidal radii equal.
Normalized in such a way, neutral hydrogen maps manifest one common feature ---
the associated HI gas appears to follow the tidal ``stretching'' axis
direction.

\section{The model, and discussion}

\begin{figure}[t]
\plotfiddle{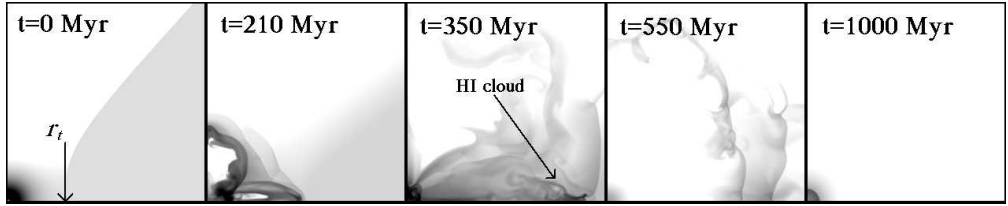}{67pt}{0}{123.6}{123.6}{-188.5}{0}
\caption{\small Results of 2D hydrodynamic simulations. The center of the dwarf galaxy
is at the lower left corner.
A dense extratidal HI cloud moving along the principal tidal axis (bottom horizontal line)
is evident for $t=210$~Myr and $t=350$~Myr.}
\label{fig2}
\end{figure}
To test the intuitive
idea that fragments of expanding HI shells resulting from supernovae type Ia explosions in dwarf spheroidals
are removed from the galaxy preferentially along the strongest tidal force direction, 2D
hydrodynamic simulations have been conducted.
The evolution of the interstellar medium (ISM) in a dSph galaxy has been followed with the ZEUS-2D hydrodynamical
code (Stone \& Norman, 1992). Figure~\ref{fig2} shows the density
distribution (logarithmic scale; black corresponds to the densest gas) obtained for a few time steps.
(One quarter of the whole distribution is shown). The box size is 5~kpc.
The axial symmetry with the axis of symmetry directed along the strongest
tidal force direction (horizontal in Figure~\ref{fig2}) has been adopted. The parameters
of the dSph model and of the tidal field correspond to the Sculptor galaxy. Initially
the hydrostatic gas with a mass $2 \times 10^5$~M$_{\odot}$ 
 and a temperature T=500~K
was perturbed by 10 supernova type Ia explosions  occuring in the center of the galaxy
at random moments of time over the interval of 1~Gyr. The assumed spatially
distributed mass input from red giants was $1.65 \times 10^{-5}$~M$_{\odot} {\rm yr}^{-1}$.
Radiative cooling was taken into account. The energy and mass inputs from the SNe
was $10^{51}$~erg and $1.4$~M$_{\odot}$ per SN.

The simulations showed that repeated SNe Ia can completely expel the ISM from a dSph
with 
most
of the mass being removed
as relatively dense clouds along the tidal axis direction. This is in accordance
with the observed HI distribution in three Local Group dwarf spheroidals. The results obtained
emphasize the impact of the tidal field 
on the evolution of
dwarf spheroidal galaxies (even for relatively ``isolated'' systems such as Tucana and Cetus),
and might shed some more light upon the complex star formation history of dSphs.




\end{document}